\documentclass[twocolumn,superscriptaddress,prl]{revtex4}
\usepackage{graphicx} 
\usepackage{amsmath,amssymb}

\begin{document}
\title{Negative entropy in scrambling black holes}

\author{Koji Azuma}
\email{koji.azuma@ntt.com}
\affiliation{Basic Research Laboratories, NTT,~Inc., 3-1 Morinosato Wakamiya, Atsugi, Kanagawa 243-0198, Japan}
\affiliation{NTT Research Center for Theoretical Quantum Information, NTT,~Inc., 3-1 Morinosato-Wakamiya, Atsugi, Kanagawa 243-0198, Japan}

\date{\today}

\begin{abstract}
We present a microscopic statistical-mechanical foundation for interpreting the horizon area of a scrambling black hole as coherent information, equivalently negative conditional quantum entropy, in Hawking's pair-creation picture.
We derive the entropy increase induced in a black hole when an infalling object is absorbed and scrambled into its microscopic degrees of freedom. Up to finite-reservoir corrections, this increase takes a canonical form at the Hawking temperature, regardless of the initial entropy carried by the infalling object. Applying this entropy formula to an incoming mode paired by time reversal with an outgoing Hawking radiation mode, we show that their partition-function contributions cancel in the coherent-information balance associated with the horizon area. 
The resulting area response is then determined only by the energy flux, in agreement with the black-hole first law.
\end{abstract}
\maketitle

{\it Introduction---}Black holes should have entropy. Bekenstein argued that this is necessary for black-hole physics to be compatible with the second law of thermodynamics \cite{B73,B74}. Hawking strengthened this idea by showing that black holes emit thermal radiation through pair creation near the event horizon, contrary to the classical point of view that nothing can escape from a black hole \cite{H74,H75}. 
This radiation fixes the temperature of the hole, called the Hawking temperature $T_H=\kappa_B\hbar/(2\pi k c)$, leading to the Bekenstein-Hawking relation that associates the horizon area $A_B$ of the black hole $B$ with a dimensionless entropy $S(B)$:
\begin{equation}
    \frac{c^3}{4G\hbar} {\rm d} A_B ={\rm d} S(B), \label{eq:beken}
\end{equation}
where $c$ is the speed of light, $G$ is the Newtonian constant of gravitation, $k$ is the Boltzmann constant, $\hbar:=h/(2\pi)$ with the Planck constant $h$, and $\kappa_B$ is the surface gravity of the black hole.
However, Hawking’s original pair-creation picture of the radiation leaves the information-theoretic meaning of entropy $S(B)$ subtle \cite{M09,BPZ13,M19,ASK25,P93}.

In particular, in Hawking’s pair-creation picture, an outgoing positive-frequency massless bosonic mode $H^+$ is accompanied by a negative-frequency one $H^-$ falling into the black hole as the purification partner. This reduces the mass and area of the hole, while the purification partner remains associated with degrees of freedom behind the horizon, making the entropy bookkeeping inside the hole nontrivial. Moreover, the presence of not only ordinary positive-energy matter but also these negative-frequency bosons inside the hole $B$ prevents a simple interpretation of the entropy $S(B)$ as an ordinary entropy of hidden degrees of freedom.  
Instead, a coherent-information interpretation offers a different viewpoint by retaining Hawking's pair-creation picture \cite{ASK25}: the horizon area $A_B$ is associated with coherent information \cite{SN96,DW05,HOW05,HOW07}, equivalently negative conditional quantum entropy,
\begin{equation}
    \frac{c^3}{4G\hbar} {\rm d} A_B = {\rm d} I(\bar{B}\rangle B^+)={\rm d}S(B^+) -{\rm d} S(B^-), \label{eq:marea} 
\end{equation}
where $S$ is the von Neumann entropy defined as $S(A):=S(\hat{\rho}_A):=-{\rm Tr}[\hat{\rho}_A\ln \hat{\rho}_A]$ for a density operator $\hat{\rho}_A$ and $I(\bar{B}\rangle B^+):=-S(\bar{B}|B^+):=S(B^+)-S(\bar{B}B^+)$ is the coherent information.
Coherent information $I(\bar{B}\rangle B^+)$ can be positive only in the quantum regime and is associated with one-way distillable entanglement \cite{DW05}.
In Eq.~(\ref{eq:marea}), the black hole $B$ is taken to consist of two subsystems: the negative subsystem $B^-$, composed only of negative-frequency massless scalar bosons, such as $H^-$, appearing in Hawking's pair-creation picture, and the positive subsystem $B^+$, the relative complement of $B^-$ in the hole $B$. The black hole $B$ has Hawking radiation once it forms, that is, ${\rm d}S(B^-)>0$, and
the system $\bar{B}$ is the purification partner of the black hole $B$ outside the hole, giving the final equality in Eq.~(\ref{eq:marea}).

However, two questions remain even for the coherent-information interpretation. First, if we throw a particle $P$ from outside into a Schwarzschild black hole $B$ with rest mass $M_B$, the first law \cite{BCH73} 
\begin{equation}
    {\rm d}(M_B c^2)=\frac{\kappa_B c^2}{8 \pi G} {\rm d}A_B \label{eq:first-law}
\end{equation}
implies that the area $A_B$ of the black hole can increase after the particle is incorporated into the positive subsystem $B^+$. This happens even if the particle $P$ is initially in a pure state. 
If the contribution to $S(B^+)$ in Eq.~(\ref{eq:marea}) were identified with the initial von Neumann entropy $S(P)=0$ of the particle $P$, then the pure-state assumption would give ${\rm d}S(B^+)=0$. This would fail to account for the area increase.

Second, a thermal Hawking mode $H^+$ with frequency $\omega$ carries a partition-function contribution to its entropy, in addition to the energy-over-temperature term (see, e.g., Ref.~\cite{ASK25}):
\begin{equation}
    S(H^+) = \frac{\hbar \omega n_{H^+}}{kT_H} +\ln Z_{H^+}(T_H),  \label{eq:radent}
\end{equation}
where $\hbar \omega n_{H^+}$ is the radiation energy at infinity and $Z_{H^+}(T_H)$ is the canonical partition function at the Hawking temperature $T_H$. 
This partition-function contribution is generally not negligible.
Since $S(H^+)=S(H^-)$ for a Hawking pair in a pure state, the negative subsystem $B^-$ that absorbs the negative-frequency bosons $H^-$ with entropy $S(H^-)$ receives the contribution ${\rm d}S(B^-)=S(H^-)=S(H^+)$. If this contribution is inserted directly into the modified area law (\ref{eq:marea}), the partition-function term appears to contribute to the horizon-area response. 
Nevertheless, the first law (\ref{eq:first-law})
predicts an area response determined only by the energy flux, i.e., ${\rm d}(M_B c^2)=-\hbar \omega n_{H^+}$ due to the outgoing Hawking radiation mode, without any contribution from the partition function.

In this paper, we show that both questions about the modified area law (\ref{eq:marea}) are resolved by microcanonical state counting in a scrambling black hole. In particular, we assume that an infalling object from outside is, after absorption, embedded into the microscopic degrees of freedom of the positive subsystem $B^+$ by scrambling. 
Up to finite-reservoir corrections, the entropy induced in the positive subsystem $B^+$ by this process has a canonical form at the Hawking temperature.
This contribution is independent of the initial entropy carried by the infalling object and thus exists even when the object is initially in a pure state.
Applying the entropy formula to an incoming massless bosonic mode paired by time reversal with an outgoing Hawking radiation mode,
the partition-function contributions cancel in the coherent-information balance, and the 
remaining area response is determined only by the energy change and agrees with the black-hole first law. 
As a result, even when the area change is evaluated microscopically from
the modified area law~(\ref{eq:marea}), the general-relativistic
first-law response is recovered not because the partition-function
contribution is neglected, but because it cancels between matched Hawking
channels after scrambling.

{\it Scrambling black hole---}Let us start by considering a scrambling Schwarzschild black hole $B$ with mass $M_B$. The first law for this hole is written as Eq.~(\ref{eq:first-law}).
The subsystem $B^\pm$ of the hole $B$ is associated with Hilbert space ${\cal H}_{B^\pm}$, and its energy is written as $E_{B^\pm}$, satisfying
\begin{equation}
    M_B c^2 = E_{B^+} +E_{B^-} \label{eq:energy}
\end{equation}
with $E_{B^+}\ge 0$ and $E_{B^-}\le 0$.

Suppose that the positive subsystem $B^+$ undergoes scrambling while imposing conservation of energy $E_{B^+}$.
This means that the positive subsystem $B^+$ relaxes to a microcanonical ensemble, that is, to the equiprobable state in the eigenspace ${\cal E}_{B^+} (E_{B^+})\subset {\cal H}_{B^+}$ associated with eigenvalue $E_{B^+}$ of the free Hamiltonian of the subsystem $B^+$, 
\begin{equation}
    \hat{\tau}_{B^+} (E_{B^+})= \frac{\hat{1}_{{\cal E}_{B^+} (E_{B^+})}}{W_{B^+}(E_{B^+})},
\end{equation}
where $\hat{1}_{\cal X}$ is the projection operator on Hilbert subspace ${\cal X}$ and $W_{X}(E_X):=\dim [{\cal E}_{X} (E_X)]$ for a system $X$.
Since the eigenvalue $E_{B^+}$ may depend on the size of the black hole, degeneracy $W_{B^+}(E_{B^+})$ is given for fixed horizon area $A_B$.
Note that the von Neumann entropy for the system $B^+$ in the maximally mixed state $\hat{\tau}_{B^+} (E_{B^+})$ on the eigenspace ${\cal E}_{B^+} (E_{B^+})$ is equivalent to the Boltzmann entropy:
\begin{equation}
S(B^+)=S(\hat{\tau}_{B^+} (E_{B^+}))=\ln W_{B^+}(E_{B^+})=:s_{B^+}(E_{B^+}).
\end{equation}

We consider an elementary process in which a small object $P$ with energy-at-infinity $E_P(\ge 0)$ is thrown from the outside into the black hole. According to the rule in Ref.~\cite{ASK25}, after crossing the event horizon, the positive part $B'^+$ of the hole is associated with the composite Hilbert space ${\cal H}_{B'^+}={\cal H}_{B^+}\otimes {\cal H}_P$ and its energy is $E_{B'^+}=E_{B^+}+E_P$. For the given $E_{B'^+}$ and $M_{B'}c^2:=M_Bc^2+E_P$, after total-energy-preserving scrambling within the positive subsystem $B'^+$, the state of the subsystem $B'^+$ can be regarded as
\begin{equation}
    \hat{\tau}_{B'^+} (E_{B'^+}) = \frac{\hat{1}_{{\cal E}_{B'^+} (E_{B'^+})}}{W_{B'^+}(E_{B'^+})} \label{eq:mixed}
\end{equation}
approximately,
where
\begin{equation}
    W_{B'^+}(E_{B'^+}) = \sum_{\epsilon_P =0}^{E_{B'^+}} W_{B^+}(E_{B'^+}-\epsilon_P) W_P(\epsilon_P). \label{eq:W}
\end{equation}
Note that the sum in this equation is the dimension of the fixed total-energy shell
written in the pre-scrambling labels of $B^+$ and $P$. The subsequent scrambling is modeled by unitaries commuting with the total Hamiltonian,
thereby randomizing states within this shell without changing its dimension.
The approximation in Eq.~(\ref{eq:mixed}) reflects the fact that both the interaction energy and the entropy associated with the initial energy distribution of $P$ are coarse-grained into the finite width of the energy shell 
\footnote{The unitaries $\hat{U}_{B'^+}$ for scrambling take the block-diagonal form $\hat{U}_{B'^+} =\bigoplus_{E_{B'^+}}\hat{U}_{{\cal E}_{B'^+}(E_{B'^+})}$. 
Then, the state after scrambling is $\bigoplus_{e_P} t(e_P)\hat{\tau}_{B'^+} (E_{B^+}+e_P)$ with the probability distribution $t(e_P) := {\rm Tr} [\hat{1}_{{\cal E}_P (e_P)} \hat{\rho}_P]$ over initial energy $e_P$, where $\hat{\rho}_P$ is the density operator of the system $P$ at the event of crossing the horizon. The von Neumann entropy of this state is thus $H(\{t(e_P)\}) + \sum_{e_P} t(e_P) \ln W_{B'^+}(E_{B^+} +e_P)$, where $H(\{t(e_P)\})$ denotes the Shannon entropy. This clarifies the assumption used in the main text that the first term $H(\{t(e_P)\})$ is negligible compared with the second term, while $\sum_{e_P} t(e_P) \ln W_{B'^+}(E_{B^+} +e_P)\approx \ln W_{B'^+}(E_{B^+} +E_P)=\ln W_{B'^+}(E_{B'^+})$ with $E_P:=\sum_{e_P} t(e_P) e_P$, reducing to the von Neumann entropy of the state in Eq.~(\ref{eq:mixed}).
}.

{\it Increase in microcanonical entropy in positive subsystem $B^+$---}To determine the response of the positive subsystem $B^+$ by absorbing a positive object $P$, we expand $s_{B^+}(E_{B'^+}-\epsilon_P)= \ln W_{B^+}(E_{B'^+}-\epsilon_P)$ in Eq.~(\ref{eq:W}) around $E_{B^+}$:
\begin{multline}
    s_{B^+}(E_{B'^+}-\epsilon_P) = s_{B^+}(E_{B^+}) +  \frac{E_P-\epsilon_P}{kT_{B^+}(E_{B^+})} \\
    + \delta_{B^+}(\epsilon_P;E_{B^+}), \label{eq:taylor}
\end{multline}
with
\begin{align}
    s'_{B^+} (E_{B^+})=:&\frac{1}{kT_{B^+}(E_{B^+})}, \label{eq:T_{B^+}}  \\
    \delta_{B^+}(\epsilon_P;E_{B^+}):=&\frac{1}{2!}s''_{B^+}(E_{B^+})(E_P-\epsilon_P)^2 \nonumber \\
    &+\frac{1}{3!} s'''_{B^+}(E_{B^+})(E_P-\epsilon_P)^3 +\cdots,
\end{align}
where $'$ denotes the partial derivative with respect to the energy at fixed $A_B$.
Since this implies
\begin{multline}
     W_{B^+}(E_{B'^+}-\epsilon_P)W_P(\epsilon_P)=W_{B^+}(E_{B^+}) e^{E_P/(kT_{B^+})} \\  \times W_P(\epsilon_P)  e^{-\epsilon_P/(kT_{B^+})} 
     e^{\delta_{B^+}},
\end{multline}
Eq.~(\ref{eq:W}) is reduced to
\begin{multline}
    \ln W_{B'^+}(E_{B'^+}) \approx \ln W_{B^+}(E_{B^+})+\frac{E_P}{kT_{B^+}}   \\
    +\ln Z_P(T_{B^+}) + \ln \langle e^{\delta_{B^+}}\rangle_{p_{T_{B^+}}}, \label{eq:approx-f}
\end{multline}
where 
\begin{equation}
    p_T(\epsilon_P)=\frac{W_P(\epsilon_P) e^{-\epsilon_P/(kT)}}{Z_P(T)} 
\end{equation}
is the canonical distribution at a temperature $T$, $Z_P(T):=\sum_{\epsilon_P (\ge 0)}W_P(\epsilon_P)e^{-\epsilon_P/(kT)}$ is the partition function,
 and $\langle f(\epsilon_P) \rangle_p$ denotes the expectation value of a function $f(\epsilon_P)$ over a probability distribution $p$. 
Hence, the increase in microcanonical entropy of the positive subsystem $B^+$ is 
\begin{multline}
    \Delta s_{B^+}:= \ln W_{B'^+}(E_{B'^+})-\ln W_{B^+}(E_{B^+})  \\
    \approx   \frac{E_P}{kT_{B^+}}
     + \ln Z_P(T_{B^+}) 
     + \ln \langle e^{\delta_{B^+}}\rangle_{p_{T_{B^+}}}.\label{eq:deltaSB^+}
\end{multline} 
This is the coarse-grained entropy increase generated by scrambling
the absorbed system into the microcanonical degrees of freedom of
\(B^+\). If \(S(B)\) in Eq.~(\ref{eq:beken}) is
interpreted as a scrambling-induced coarse-grained entropy, the
corresponding formula for the ordinary area law is obtained from
Eq.~(\ref{eq:deltaSB^+}) by replacing \(T_{B^+}\) with the Hawking temperature \(T_H\).

{\it Cavity calibration of the temperature $T_{B^+}$---}We next calibrate the temperature $T_{B^+}$ of the positive subsystem $B^+$ defined in Eq.~(\ref{eq:T_{B^+}}). This temperature is not determined by the state-counting identity itself. It is a parameter characterizing the local slope of the density of states of the positive subsystem at fixed horizon area $A_B$.

Consider placing the black hole in an ideal reflecting cavity. An outgoing Hawking mode is then returned to the black hole as the corresponding time-reversed incoming channel. The cavity is used only as a reference calibration: it is not meant to describe the evaporation process itself. In this reference process, the outgoing Hawking channel and its time-reversed absorption channel form a stationary fixed-area exchange.

We choose the incoming channel $P$ to be the time-reversed counterpart of the outgoing Hawking mode $H^+$. Thus the incoming oscillator mode $P$ has the same energy spectrum as the outgoing mode $H^+$, and the energy assigned to the incoming channel in the reference process is the mean Hawking-channel energy,
$
    E_P=\hbar\omega n_{H^+}.
$    
At fixed horizon area, Eq.~(\ref{eq:marea}) gives
\(
    \Delta s_{B^+}=\Delta S(B^-).
\)
On the other hand, the contribution to the negative subsystem $B^-$ is supplied by the negative-frequency partner $H^-$ of the Hawking pair in a pure state as
\(
    \Delta S(B^-)=S(H^-)=S(H^+).
\)
Using Eq.~(\ref{eq:deltaSB^+}) for the positive-sector contribution and Eq.~(\ref{eq:radent}) for the Hawking contribution, the reference balance becomes
\begin{multline}
    \frac{\hbar\omega n_{H^+}}{kT_{B^+}}
    +\ln Z_P(T_{B^+})
    +\ln\langle e^{\delta_{B^+}}\rangle_{p_{T_{B^+}}} \\
    =
    \frac{\hbar\omega n_{H^+}}{kT_H}
    +\ln Z_{H^+}(T_H).
\end{multline}
Since the reference incoming channel $P$ is the time reverse of the Hawking mode, it has the same energy spectrum. 
Hence, in the leading reservoir approximation, the reference balance is reproduced by assigning the same temperature to the absorption channel,
\begin{equation}
    T_{B^+}=T_H . \label{eq:temp}
\end{equation}

Once the positive-sector temperature $T_{B^+}$ has been calibrated to the Hawking temperature, Eq.~(\ref{eq:deltaSB^+}) can be applied to a general incoming channel. The partition-function term then arises from the positive-sector state count itself; it is not imposed by the calibration and is not discarded.

{\it Positive-sector entropy induced by scrambling---}Combining Eq.~(\ref{eq:deltaSB^+}) with the calibration in Eq.~(\ref{eq:temp}), we obtain
\begin{equation}
\Delta s_{B^+}
\approx
\frac{E_P}{kT_H}
+\ln Z_P(T_H)
+\ln\langle e^{\delta_{B^+}}\rangle_{p_{T_H}} .
\end{equation}
In the leading reservoir approximation, where the last term is negligible, this becomes
\begin{equation}
\Delta s_{B^+}
\approx
\frac{E_P}{kT_H}
+\ln Z_P(T_H). \label{eq:main}
\end{equation}
This increase has a canonical form at the Hawking temperature $T_H$, but it is not the entropy of an initially thermal state of \(P\): \(E_P\) is the energy at infinity carried by the infalling object, whereas \(Z_P(T_H)\) arises from the post-absorption state counting. 
Thus, \(\Delta s_{B^+}\) is determined by \(E_P\) and the spectrum of \(P\), rather than by the entropy initially carried by \(P\); in particular, even when the object \(P\) is initially in a pure state, its absorption and subsequent scrambling induce the entropy increase \(\Delta s_{B^+}\).

{\it Mode-by-mode cancellation of partition-function contributions in Hawking processes---}We now include Hawking radiation using the leading-order expression (\ref{eq:main}) for $\Delta s_{B^+}$.
For each outgoing Hawking mode with positive frequency $\omega$, one can define a corresponding incoming scattering mode with the same frequency and quantum numbers, equivalently the time-reversed 
channel of the outgoing radiation mode. We then consider a process in which a Hawking pair $H^+H^-$ is produced while a massless scalar boson $P$ with energy-at-infinity $E_P$ in the corresponding incoming channel enters the black hole. The incoming mode $P$ has the same frequency $\omega$ as the Hawking radiation mode $H^+$.

As the Hawking pair \(H^+H^-\) is in a pure state, the entropy increase $\Delta S(B^-)$ of the negative subsystem $B^-$ is $\Delta S(B^-)=S(H^-) =S(H^+)$. 
The entropy \(S(H^+)\) is given by Eq. (4), with \(n_{H^+}=(e^{\hbar\omega/(kT_H)}-1)^{-1}\) and \(Z_{H^+}(T_H)=(1-e^{-\hbar\omega/(kT_H)})^{-1}\) (see, e.g., Ref.~\cite{ASK25}).
Since bosonic modes $P$ and $H^+$ have the same frequency $\omega$, we have \(Z_P(T_H)=Z_{H^+}(T_H)\), and the partition-function contributions cancel in the coherent-information balance:
\begin{equation}
    \Delta s_{B^+} -\Delta S(B^-) \approx  \frac{E_P}{kT_H} -\frac{\hbar \omega n_{H^+}}{kT_H} 
    = \frac{\Delta (M_Bc^2)}{kT_H}, \label{eq:balance}
\end{equation}
where we used Eq.~(\ref{eq:main}) and the variation of Eq.~(\ref{eq:energy}), $\Delta(M_Bc^2)=\Delta E_{B^+}+\Delta E_{B^-} =E_P-\hbar \omega n_{H^+}$, in the final equality. 
Combined with the assumed modified area law (\ref{eq:marea}) 
identifying the positive-sector contribution \({\rm d}S(B^+)\) with
\(\Delta s_{B^+}\), this reduces to the first law (\ref{eq:first-law}) on a mode-by-mode basis.
That is, provided that outgoing Hawking channels and incoming massless bosonic channels can be paired by time reversal,
the area change $\Delta A_B$ is characterized solely by the energy change, as predicted by the first law (\ref{eq:first-law}). 

Such a cancellation of partition-function contributions has no counterpart in the ordinary area-law
interpretation of Eq.~(\ref{eq:beken}) based on scrambling-induced
coarse graining, where the absorption-induced entropy contributes
directly to \(S(B)\) rather than to a sector-resolved balance
involving the negative Hawking partner \(H^-\). 
This implies that the area response in our model with Hawking radiation differs from that in such a scrambling-induced
coarse-graining black hole without Hawking radiation.

In the leading reservoir approximation,
the only ingredient needed to reduce the modified area law (\ref{eq:marea}) to the first law in Eq.~(\ref{eq:first-law}) is the equality of the partition functions of the outgoing Hawking mode $H^+$
 and its time-reversed incoming mode $P$.
This is in striking contrast to the derivation in Ref.~\cite{ASK25}, where the system $P$ had to be in the same Gibbs state as the Hawking particles $H^+$ before falling in, and only quasi-static processes were treated: in our case, the reduction is achieved regardless of the initial quantum state of the incoming boson $P$, even if the massless boson $P$ is initially in a pure state, and processes beyond a quasi-static process can be treated as long as scrambling in the hole is fast \cite{SS08}.

{\it Discussion---}If the positive subsystem $B^+$ represents the ordinary part of a black hole, it may be natural to model it by microcanonical state counting and scrambling \cite{SS08,SV96,HP07,H90,NWK23}. 
Here, however, they play a different role: rather than merely counting hidden black-hole degeneracies, they provide a microscopic statistical-mechanical realization of the coherent-information interpretation of black-hole area. The central idea is that the entropy relevant to the area response is not the initial von Neumann entropy of the infalling object. After absorption, scrambling embeds the object into the black-hole microcanonical degrees of freedom, and the state-count ratio induces the canonical entropy contribution in Eq.~(\ref{eq:main}). This explains why the area response is obtained even when the infalling object is initially in a pure state.

This mechanism also clarifies why the partition-function contribution in the entropy of a Hawking mode does not appear in the area response predicted by the first law (\ref{eq:first-law}). Although a thermal Hawking mode contains such a contribution, the coherent-information balance removes it through cancellation between massless bosonic channels matched by time reversal. As a result, only the energy imbalance remains in Eq.~(\ref{eq:balance}), and the first-law response is recovered.

The cancellation relies on the existence of matched incoming and outgoing channels. It is therefore most directly applicable to massless, or sufficiently light, incoming bosonic fields with corresponding Hawking radiation channels. For ordinary massive incoming elementary-particle modes, even when no
corresponding Hawking channel is included, the partition-function
contribution is Boltzmann
suppressed at the Hawking temperature, provided no large degeneracy
compensates this suppression; in such sectors the area response is
therefore well approximated by the first-law form. 
Hence, except for atypical inputs, the area response given by Eq.~(\ref{eq:marea}) agrees with the first-law response in Eq.~(\ref{eq:first-law}) to leading order.

For general stationary black holes, the same microcanonical reasoning applies at first order once the energy is measured relative to the rotating and charged horizon; the induced canonical weight is then replaced by a grand-canonical one, as shown in Appendix A. With the same replacement, the mode-by-mode cancellation of partition-function contributions also extends to matched Hawking channels, as summarized in Appendix B. Finite-reservoir corrections and modes outside the positive-effective-frequency sector require separate treatment.

The coherent-information interpretation of black-hole area also points
to a broader question: what statistical mechanics underlies the area term
in Hawking's pair-creation picture? In the bookkeeping developed here,
the positive sector is described by microcanonical state counting after
scrambling, whereas the negative sector is described by the von Neumann
entropy of the purification partners produced in Hawking pair creation.
Thus the horizon-area response is not associated with an ordinary entropy
of a single hidden system, but with a sector-resolved balance between a
microcanonical contribution from \(B^+\) and a von Neumann entropic
contribution from \(B^-\). 
A more explicit microscopic realization of this balance, analogous in spirit to the Strominger--Vafa state counting of black-hole entropy \cite{SV96}, would further sharpen the statistical-mechanical meaning of the area term in Hawking's pair-creation picture.

\begin{acknowledgments}
{\it Acknowledgments---}I thank R. Jozsa for raising the question of what happens if a black hole receives a system in a pure state. I also thank G.~Kato, K.~Sawada, S.~Subramanian, K.~Tamaki, Y.~Yamada, and T.~Yamazaki for helpful discussions. This work was supported in part by Moonshot R\&D, JST JPMJMS256E, and by R\&D of ICT Priority Technology (JPMI00316).
\end{acknowledgments}
{\it Data availability---}All the data supporting the findings of this study are available within the paper.
\appendix
\setcounter{equation}{0}
\renewcommand{\theequation}{A\arabic{equation}}

{\it Appendix A: Entropy increase in general scrambling black holes}---We briefly describe how the microcanonical state-counting argument extends to a general stationary black hole $B$. The first law for the black hole is written as
\begin{equation}
    {\rm d}(M_Bc^2) = \frac{\kappa_Bc^2}{8\pi G} {\rm d}A_B + \Omega_B {\rm d}J_B+\Phi_B {\rm d}Q_B,  \label{eq:firstlawg}
\end{equation}
where $J_B$ is the angular momentum, $\Omega_B$ is the angular velocity, $Q_B$ is the charge, and $\Phi_B$ is the electrostatic potential of the black hole (see, e.g., Ref.~\cite{H76} for their explicit forms).

For stationary black holes, the sector decomposition should be understood
with respect to the history of the hole. We take the negative subsystem
$B^-$ to be empty at the moment of black-hole formation. It is
populated only by the negative-effective-energy partners $H^-$ generated in
Hawking's pair-creation processes \cite{ASK25}. Ordinary matter $P$ absorbed from outside
is assigned to the positive subsystem \(B^+\). In particular, by
Hawking's area theorem \cite{H72}, classical matter crossing the horizon has
nonnegative horizon-generator energy
\[
K_P:=E_P-\Omega_B L_P-\Phi_B Q_P\ge 0,
\]
called ``effective energy'' in what follows,
where \(E_P\), \(L_P\), and \(Q_P\) are, respectively, the energy at infinity,
the axial component of angular momentum, and the charge of the object,
evaluated at the event of crossing \cite{MTW}. This quantity $K_P$ was referred to as a ``kinetic energy'' in Ref.~\cite{ASK25}.
Such ordinary matter therefore belongs to the positive
sector $B^+$ and cannot be confused with the negative sector \(B^-\).
Thus, in the absorption process considered in the state-counting
argument, the incoming object changes only the positive-sector charges,
whereas the negative sector is changed only by Hawking-pair creation with negative-frequency particles $H^-$ having a negative-effective energy $K_{H^-}<0$.

Accordingly, we decompose the globally conserved quantities into
sector-resolved bookkeeping variables,
\begin{align}
&X_{B^\pm}:=(X_{B^\pm}^i):=(E_{B^\pm},J_{B^\pm},Q_{B^\pm}), \\
&X_B=X_{B^+}+X_{B^-}.
\end{align}
For an ordinary absorbed object $P$ with 
\[
Y_P:=(Y_P^i):=(E_P,L_P,Q_P),
\]
the absorption step acts only on the positive sector,
\begin{align}
X_{B^{\prime +}}&=X_{B^+}+Y_P,
\end{align}
consistent with the conservation law \cite{MTW} for energy, angular momentum, and charge.
The subsequent scrambling is assumed to act within the positive sector $B'^+$ while preserving the total sector-resolved charges $X_{B^{\prime +}}$.
The negative sector is enlarged separately when a Hawking pair is
created.

Let us consider the entropy increase $\Delta s_{B^+}$ by throwing a small positive-effective-energy particle $P$ into the black hole $B$.
Throughout this state-counting step, the density of states of \(B^+\)
is understood as a conditional density of states at fixed \(X_{B}\),
\(W_{B^+}(X_{B^+};X_{B})\). For simplicity, we suppress this dependence below.
For given $X_{B'^+}$ and $X_{B'}:=X_B+Y_P$, after $X_{B'^+}$-preserving scrambling on the positive subsystem $B'^+$, 
the degeneracy \(W_{B^{\prime +}}(X_{B'^+})\) of the positive subsystem \(B^{\prime +}\)
can be written in terms of the degeneracies of the original systems
\(B^+\) and \(P\) as
\begin{equation}
    W_{B'^+}(X_{B'^+}) = \sum_{y_P} W_{B^+}(X_{B'^+}-y_P) W_P(y_P) \label{eq:deg}
\end{equation}
approximately, where 
\[y_P=(\epsilon_P, l_P,q_P).\]
By applying the Taylor expansion to
\(s_{B^+}(X_{B^{\prime +}}-y_P;X_{B})
:=\ln W_{B^+}(X_{B^{\prime +}}-y_P;X_{B})\)
around \(X_{B^+}\), we have
\begin{multline}
    s_{B^+}(X_{B'^+}-y_P) = s_{B^+}(X_{B^+}) +   \frac{E_P- \epsilon_P}{k T_{B^+}} \\- \frac{\Omega_{B^+} (L_P-l_P)}{k T_{B^+}}-\frac{\Phi_{B^+}(Q_P-q_P)}{k T_{B^+}}   
    +  \delta_{B^+}(y_P;X_{B^+}) , \label{eq:taylorg}
\end{multline}
where 
\begin{align}
    &\frac{\partial s_{B^+}(X_{B^+})}{\partial X_{B^+}^1}=:\frac{1}{kT_{B^+}(X_{B^+})} ,\label{eq:temp-g}\\
    & \frac{\partial s_{B^+}(X_{B^+})}{\partial X_{B^+}^2}=:-\frac{\Omega_{B^+}(X_{B^+})}{kT_{B^+}(X_{B^+})},\\
    & \frac{\partial s_{B^+}(X_{B^+})}{\partial X_{B^+}^3}=:-\frac{\Phi_{B^+}(X_{B^+})}{kT_{B^+}(X_{B^+})} ,\label{eq:phi-g}
\end{align}
with partial derivatives with respect to
\(X_{B^+}\) at fixed \(X_{B}\),
and
\begin{multline}
    \delta_{B^+}(y_P;X_{B^+})\\
    :=\frac{1}{2!}\sum_{i,j} \frac{\partial^2 s_{B^+}(X_{B^+})}{\partial X_{B^+}^i \partial X_{B^+}^j} (Y_P^i-y_P^i) (Y_P^j-y_P^j)   \\
    + \frac{1}{3!}\sum_{i,j,k} \frac{\partial^3 s_{B^+}(X_{B^+})}{\partial X_{B^+}^i \partial X_{B^+}^j \partial X_{B^+}^k} (Y_P^i-y_P^i)(Y_P^j- y_P^j)(Y_P^k -y_P^k)  \\+
    \cdots.
\end{multline}
Since this implies 
\begin{multline}
     W_{B^+}(X_{B'^+}-y_P)W_P(y_P)=W_{B^+}(X_{B^+}) \\ 
     \times e^{(E_P-\Omega_{B^+}L_P-\Phi_{B^+}Q_P)/(kT_{B^+})} \\  \times W_P(y_P)  e^{-(\epsilon_P-\Omega_{B^+}l_P-\Phi_{B^+}q_P)/(kT_{B^+})} 
     e^{\delta_{B^+}},
\end{multline}
Eq.~(\ref{eq:deg}) is reduced to
\begin{multline}
    \ln W_{B'^+}(X_{B'^+}) \approx \ln W_{B^+}(X_{B^+}) \\
    +\frac{E_P-\Omega_{B^+}L_P-\Phi_{B^+}Q_P}{kT_{B^+}} \\
    +\ln \Xi_P(\chi_{B^+}) + \ln \langle e^{\delta_{B^+}}\rangle_{r_{\chi_{B^+}}}, \label{eq:approx-f-g}
\end{multline}
where 
\begin{equation}
    \Xi_P(\chi_{B^+}):=\sum_{y_P} W_P(y_P) e^{-(\epsilon_P-\Omega_{B^+}l_P-\Phi_{B^+}q_P)/(kT_{B^+})}
\end{equation}
is the grand-canonical partition function at 
\[\chi_{B^+}:=(T_{B^+},\Omega_{B^+},\Phi_{B^+}),
\]
and
\begin{equation}
    r_{\chi_{B^+}}(y_P)=\frac{W_P(y_P) e^{-(\epsilon_P-\Omega_{B^+}l_P-\Phi_{B^+}q_P)/(kT_{B^+})}}{\Xi_P(\chi_{B^+})}  
\end{equation}
is the grand-canonical distribution.
Hence, we have
\begin{multline}
    \Delta s_{B^+}= s_{B'^+}(X_{B'^+})-s_{B^+}(X_{B^+})  \\
    \approx \frac{E_P-\Omega_{B^+}L_P-\Phi_{B^+}Q_P}{kT_{B^+}} \\
     + \ln \Xi_P(\chi_{B^+})+ \ln \langle e^{\delta_{B^+}}\rangle_{r_{\chi_{B^+}}}.\label{eq:deltaSB^+g}
\end{multline}

As in the case for Schwarzschild black holes, we can calibrate conjugate parameters $T_{B^+}$, $\Omega_{B^+ }$, and $\Phi_{B^+}$ in Eqs.~(\ref{eq:temp-g})-(\ref{eq:phi-g})---defined by the local derivatives of \(s_{B^+}(X_{B^+})\), not
independent geometric properties of the positive sector---as
\begin{equation}
    \chi_{B^+} = \chi_{B}:=(T_H, \Omega_B,\Phi_B),\label{eq:chi}
\end{equation} 
by using an ideal cavity as a reference process in which a Hawking
emission channel and its time-reversed absorption channel form, at the
level of the leading reservoir approximation, a stationary exchange with
no net change in the total black-hole bookkeeping variables \(X_B\), and
hence no net area change in Eq.~(\ref{eq:marea}).
The relevant Hawking-channel entropy entering this reference balance is written \cite{H76,P76} as
\begin{equation}
    S(H^+)=\frac{\hbar\omega'n_{H^+}}{kT_H} +\ln \Xi_{H^+}(\chi_B) \label{eq:radg}
\end{equation}
where $n_{H^+}=(e^{\hbar \omega' / (kT_H)}-1)^{-1}$ is the average occupation number, $\Xi_{H^+}(\chi_B)=(1-e^{-\hbar \omega'/(kT_H)})^{-1}$ is the partition function, 
$\omega':=\omega-m \Omega_B-e \Phi_B/\hbar(>0)$ is an effective frequency with the mode frequency $\omega (>0)$, the angular momentum $m\hbar$ about the axis of rotation of the black hole, and the charge $e$ carried by each quantum of the mode (see, e.g., Ref.~\cite{ASK25}).

Therefore, with Eq.~(\ref{eq:chi}), Eq.~(\ref{eq:deltaSB^+g}) reduces to
\begin{equation}
    \Delta s_{B^+}
    \approx \frac{K_P}{kT_H} 
     + \ln \Xi_P(\chi_{B})+ \ln \langle e^{\delta_{B^+}}\rangle_{r_{\chi_{B}}}.\label{eq:entropyinc}
\end{equation}
The last term is the finite-reservoir correction; the calibration in
Eq.~(\ref{eq:chi}) fixes only the linear conjugate parameters and does not remove
this correction.

\setcounter{equation}{0}
\renewcommand{\theequation}{B\arabic{equation}}

{\it Appendix B: Mode-by-mode cancellation of partition-function contributions in Hawking processes for general black holes---}Let us consider the Hawking radiation from a general black hole, whose entropy is written as Eq.~(\ref{eq:radg}).
As in the case for Schwarzschild black holes, suppose that a Hawking pair is produced while a massless scalar boson $P$ in the time-reversed scattering mode with the same frequency $\omega$, angular momentum $m\hbar$ and charge $e$ falls into the black hole. 
For this process, \(\Delta S(B^-)=S(H^-)=S(H^+)\) with $S(H^+)$ given by Eq.~(\ref{eq:radg}), and Eq.~(\ref{eq:entropyinc}) gives the
positive-sector contribution. Since the modes $P$ and $H^+$ matched by time reversal have the same energy spectrum, the partition-function contributions in the coherent-information balance cancel:
\begin{multline}
    \Delta s_{B^+}-\Delta S(B^-)\approx \frac{K_P}{kT_H} -\frac{\hbar\omega'n_{H^+}}{kT_H}  \\
    =\frac{E_P-\hbar \omega n_{H^+}}{kT_H} -\frac{\Omega_B (L_P-m\hbar n_{H^+})}{kT_H}\\-\frac{\Phi_B(Q_P-en_{H^+})}{{kT_H}} \\
    =\frac{\Delta (M_Bc^2)-\Omega_B \Delta J_B -\Phi_B \Delta Q_B}{kT_H}
\end{multline}
up to finite-reservoir correction, 
where we used the conservation law \cite{MTW} for the black hole. 
Combined with the modified area law (\ref{eq:marea}) identifying the positive-sector contribution \({\rm d} S(B^+)\) with
\(\Delta s_{B^+}\), this reduces to the first law (\ref{eq:firstlawg}).


\begin{thebibliography}{99}
    \bibitem{B73}
    J.~D.~Bekenstein, Black holes and entropy, Phys.~Rev. D {\bf 7}, 2333 (1973).
    \bibitem{B74}
    J.~D.~Bekenstein,
    Generalized second law of thermodynamics in black hole physics, Phys. Rev. D {\bf 9}, 3292 (1974).
    \bibitem{H74}
    S.~W.~Hawking, Black hole explosions?,
    Nature {\bf 238}, 30 (1974).
    \bibitem{H75}
    S.~W.~Hawking,
    Particle creation by black holes,
    Commun. Math. Phys. {\bf 43}, 199 (1975).
    \bibitem{P93}
    D. N. Page, Information in black hole radiation, Phys. Rev. Lett. {\bf 71}, 3743 (1993).
    \bibitem{M09}
    S.~D.~Mathur,
    The information paradox: a pedagogical introduction,
    Class. Quant. Grav. {\bf 26}, 224001 (2009).
    \bibitem{BPZ13} S.~L.~Braunstein, S.~Pirandola, and K.~\.{Z}yczkowski,
    Better late than never: Information retrieval from black holes,
    Phys. Rev. Lett. {\bf 110}, 101301 (2013).
    \bibitem{M19}
    S.~D.~Mathur, 
    Resolving the black hole causality paradox,
    General Relativity and Gravitation {\bf 51}, 24 (2019).
    \bibitem{ASK25}
    K.~Azuma, S. Subramanian, and G. Kato, Do black holes store negative entropy?, 
    Prog. Theor. Exp. Phys. {\bf 2025}, 053A01 (2025).
    \bibitem{SN96}
    B.~Schumacher and M.~A.~Nielsen,
    Quantum data processing and error correction,
    Phys.~Rev.~A {\bf 54}, 2629 (1996).
    \bibitem{DW05}
    I.~Devetak and A.~Winter,
    Distillation of secret key and entanglement from quantum states,
    Proc. R. Soc. Lond. A {\bf 461}, 207 (2005).
    \bibitem{HOW05}
    M.~Horodecki, J.~Oppenheim, and A.~Winter,
    Partial quantum information,
    Nature {\bf 436}, 673 (2005).
    \bibitem{HOW07}
    M.~Horodecki, J.~Oppenheim, and A.~Winter,
    Quantum state merging and negative information,
    Commun.~Math.~Phys. {\bf 269}, 107 (2007).
    \bibitem{BCH73}
    J.~M.~Bardeen, B.~Carter, and S.~W. Hawking,
    The four laws of black hole mechanics, 
    Commun. Math. Phys. {\bf 31}, 161 (1973).
    \bibitem{SS08}
    Y. Sekino and L. Susskind, 
    Fast scramblers,
    J. High Energy Phys. {\bf 10}, 065 (2008).
    \bibitem{H90}
    G.~'t Hooft,
    On the quantum structure of a black hole, 
    Nucl. Phys. B {\bf 256}, 727 (1985).
    \bibitem{SV96}
    A.~Strominger and C.~Vafa,
    Microscopic origin of the Bekenstein-Hawking entropy,
    Phys. Lett. B {\bf 379}, 99 (1996).
    \bibitem{HP07}
    P.~Hayden and J.~Preskill, 
    Black holes as mirrors: quantum information in random subsystems,
    J. High Energy Phys. {\bf 09}, 120 (2007).
    \bibitem{NWK23}
    Y.~Nakata, E.~Wakakuwa, and M.~Koashi,
    Black holes as clouded mirrors: the Hayden-Preskill protocol with symmetry,
    Quantum {\bf 7}, 928 (2023).
    \bibitem{H76}
    S.~W.~Hawking, 
    Black holes and thermodynamics,
    Phys. Rev. D {\bf 13}, 191 (1976).
    \bibitem{H72}
    S.~W.~Hawking,
    Black holes in general relativity,
    Commun. Math. Phys. {\bf 25}, 152 (1972).
    \bibitem{MTW} 
    C. W. Misner, K. S. Thorne, and J. A. Wheeler, {\it Gravitation} (W. H. Freeman, San Francisco, 1973).
    \bibitem{P76}
    D. N. Page, Particle emission rates from a black hole. II. Massless particles from a rotating hole, Phys. Rev. D {\bf 14}, 3260 (1976).
\end{thebibliography}
\end{document}